\documentclass[runningheads]{llncs}

\usepackage[T1]{fontenc}
\usepackage{graphicx}
\usepackage{booktabs}
\usepackage{csquotes}
\usepackage{placeins}
\usepackage{adjustbox}
\usepackage{comment}
\usepackage{amsmath,amssymb}
\usepackage[hidelinks]{hyperref}
\usepackage{caption}

\begin{document}

\title{Adversarial Evasion in Non-Stationary Malware Detection: Minimizing Drift Signals through Similarity-Constrained Perturbations}
\titlerunning{Adversarial Evasion in Non-Stationary Malware Detection}

\author{Pawan Acharya \and Lan Zhang}
\authorrunning{Acharya and Zhang}

\institute{Northern Arizona University, Flagstaff, AZ, USA\\
\email{pa577@nau.edu, lan.zhang@nau.edu}}

\maketitle

\begin{abstract}
Deep learning has emerged as a powerful approach for malware detection, demonstrating impressive accuracy across various data representations. However, these models face critical limitations in real-world, non-stationary environments where both malware characteristics and detection systems continuously evolve. Our research investigates a fundamental security question: Can an attacker generate adversarial malware samples that simultaneously evade classification and remain inconspicuous to drift monitoring mechanisms?
We propose a novel approach that generates targeted adversarial examples in the classifier's standardized feature space, augmented with sophisticated similarity regularizers. By carefully constraining perturbations to maintain distributional similarity with clean malware, we create an optimization objective that balances targeted misclassification with drift signal minimization.
We quantify the effectiveness of this approach by comprehensively comparing classifier output probabilities using multiple drift metrics.
Our experiments demonstrate that  similarity constraints can reduce output drift signals, with $\ell_2$ regularization showing the most promising results. We observe that perturbation budget significantly influences the evasion-detectability trade-off, with increased budget leading to higher attack success rates and more substantial drift indicators.

\end{abstract}


\section{Introduction}
In real-world cybersecurity environments, malware detection systems face a fundamental challenge: the inherently dynamic nature of malicious software \cite{akerele2025modern}. Modern malware detection models predominantly operate on static, tabular feature vectors such as byte-frequency statistics and structural metadata which can be efficiently extracted at scale and integrated into standard classification pipelines. However, these systems are continuously challenged by the rapid evolution of malware, where attackers constantly modify their techniques through obfuscation, polymorphism, packing, and build-chain alterations \cite{zhang2022semantics}.

To address this perpetual transformation, organizations have developed non-stationary malware detection systems with sophisticated drift monitoring mechanisms. These systems aim to detect and respond to significant changes in data distributions or model performance. Classic drift detection approaches include methods like DDM and EDDM, which monitor error statistics and trigger alerts when misclassification patterns deviate from historical baselines \cite{gama2004learning,baena2006early}. When ground truth labels are delayed or unavailable, unsupervised drift monitors employ statistical techniques such as Kolmogorov-Smirnov(KS) tests, Jensen-Shannon(JS) divergence, Wasserstein distance, and Maximum Mean Discrepancy(MMD) to compare data distributions across different time windows \cite{hinder2024adversarial,yang2022concept}.

The landscape of adversarial malware generation has evolved to address detection challenges through various sophisticated approaches. Researchers have developed gradient-based attacks specifically adapted to malware models \cite{kolosnjaji2018adversarial,zhang2022semantics}, creating methods that can manipulate feature representations to evade detection. Search-based approaches utilizing evolutionary algorithms have emerged, allowing for more nuanced modifications of malware samples \cite{anderson2018learning}. Reinforcement learning frameworks have further advanced these techniques \cite{zhang2022semantics}.

Existing adversarial malware generation methods remain limited by key constraints. Most target detection models in a static setting, assuming models do not change over time. This overlooks concept drift in real-world malware detection, where malware and defenses continually co-evolve. Many approaches struggle to preserve malware semantics while achieving evasion, and they rarely evaluate whether generated adversarial samples remain similar to genuine malware. This context gives rise to a pivotal research question: In a non-stationary malware detection environment, can an attacker generate adversarial samples that not only evade classification but also remain inconspicuous to drift monitoring mechanisms? This challenge is especially acute in adversarial security, where uncoordinated distribution shifts can be deliberately exploited.

To address this  challenge, we propose a novel approach that generates targeted adversarial examples in the classifier's standardized feature space, augmented with sophisticated similarity regularizers including Kullback-Leibler (KL) divergence, $\ell_2$ distance, and MMD. By carefully constraining perturbations to maintain distributional similarity with clean malware, we create an optimization objective that balances targeted misclassification with drift signal minimization.  We quantify the effectiveness of this approach by comprehensively comparing classifier output probabilities using multiple drift metrics, including JS Divergence, Hellinger Distance, Wasserstein Metric, and others, thereby systematically probing the vulnerabilities of current concept drift detection strategies.

Our experimental findings reveal three key insights into adversarial malware generation. Under a constrained low-budget setting, different similarity constraints demonstrate varying drift mitigation capabilities, with the $\ell_2$ regularization approach effectively minimizing drift signals. The iterative Fast Gradient Sign Method (i-FGSM) with $\ell_2$ constraints lowered KS and Population Stability Index (PSI) metrics most effectively, while MMD regularization introduced substantially larger distributional shifts. 
These findings underscore the nuanced challenge of generating adversarial malware samples that can simultaneously evade detection and maintain distributional similarity, providing critical insights into the vulnerabilities of current concept drift monitoring mechanisms.

Our contributions are summarized as follows:
\begin{itemize}
\item We formulate similarity-constrained targeted \emph{feature-space} attacks for tabular malware detection by augmenting targeted cross-entropy with KL, $\ell_2$, and MMD regularizers.
\item We propose an drift evaluation protocol comparing clean vs.\ adversarial malware using JSD, Hellinger, Wasserstein, output-MMD, KS, and PSI.
\item We characterize evasion–detectability trade-offs across objectives, optimizers, and hyperparameters, showing perturbation budget as the dominant driver of ASR and drift.
\end{itemize}

\section{Background and Related Work}

\subsection{Non-stationary Malware Detection and Concept Drift}
The cybersecurity landscape is defined by the continuous evolution of malware, with attackers constantly developing new obfuscation and modification techniques \cite{guerra2022android,guerra2024experts,zhang2022semantics}. This perpetual transformation challenges traditional static machine learning models, which quickly become obsolete as malware characteristics shift.
Concept drift represents the fundamental challenge in maintaining effective malware detection systems \cite{gama2004learning,baena2006early}. It describes how statistical properties of target variables change over time, rendering previously trained models increasingly ineffective. Researchers have developed two primary strategies: retraining, which replaces existing models with new ones, and incremental algorithms that continuously update models as new information emerges \cite{li2024revisiting}.

Drift detection mechanisms have become crucial in managing this non-stationary environment. These mechanisms employ sophisticated statistical techniques to identify and flag anomalous inputs that deviate from the model's expected behavior \cite{hinder2024one}. Classic methods like Drift Detection Method (DDM) and Early Drift Detection Method (EDDM) monitor error rates and error spacing, providing early warnings about potential distributional shifts \cite{gama2004learning,baena2006early}.
The ultimate objective of non-stationary malware detection is to create adaptive systems capable of maintaining high detection performance in the face of continuous technological evolution \cite{guerra2024experts}. By implementing drift monitoring techniques and flexible model strategies, researchers aim to develop more resilient detection mechanisms that can effectively respond to ever-changing landscape of malware threats.

\subsection{Adversarial Malware Generation Methods}
Adversarial malware generation has emerged as a critical research domain, demonstrating the vulnerabilities of learning-based malware detection systems \cite{kolosnjaji2018adversarial,anderson2018learning}. Existing approaches can be categorized into three methodological families: gradient-based attacks, generative methods, and reinforcement learning techniques.

Gradient-based attacks leverage loss gradients to identify precise modifications that alter model predictions, strategically manipulating executable files to bypass detection with minimal changes. Researchers have demonstrated the effectiveness of systematically appending or modifying bytes while maintaining the executable's core functionality \cite{kolosnjaji2018adversarial,kreuk2018deceiving}. 
Generative approaches train models to produce adversarial feature vectors that can transfer across detection systems, focusing on byte-level insertion strategies that maximize evasion potential. These methods represent a more sophisticated approach to adversarial sample generation \cite{hu2022generating}. 
Reinforcement learning-based methods introduce a dynamic approach, framing adversarial malware generation as a Markov Decision Process \cite{anderson2018learning,zhang2022semantics}. RL agents learn to apply binary transformations that maximize evasion against malware detectors, using advanced optimization techniques \cite{anderson2018learning}. 

However, a critical limitation persists: most existing approaches evaluate success primarily through misclassification and executability preservation, overlooking how adversarial samples interact with drift monitoring mechanisms \cite{zhang2022semantics,guerra2024experts}. 

\section{Methodology}
\label{sec:methodology}

Our research addresses a critical challenge in non-stationary malware detection by developing adversarial techniques that can simultaneously evade classification and remain inconspicuous to drift monitoring mechanisms. We propose a novel approach generating targeted adversarial examples within the classifier's standardized feature space, focusing on static, fixed-length tabular representations derived from byte-level statistics and structural metadata. By incorporating sophisticated similarity-constrained approaches with regularization techniques based on KL divergence, $\ell_2$ distance, and MMD, we develop a nuanced framework for creating adversarial samples that can potentially bypass detection while maintaining statistical similarity to original malware distributions. Our comprehensive methodology systematically evaluates the effectiveness of these adversarial samples through a multi-dimensional analysis, employing a sophisticated suite of drift metrics including JS divergence, Hellinger distance, Wasserstein distance, output-based MMD, KS statistic, and PSI.

\subsection{Framework Overview}
\label{subsec:overview}

Let $f_\theta : \mathbb{R}^d \rightarrow [0,1]^2$ denote a trained binary classifier that outputs class probabilities for benign and malicious classes. We assume access to feature vectors $x \in \mathbb{R}^d$ corresponding to malware samples ($y=1$). Given a set of clean malware features $X = \{x_i\}_{i=1}^m$, our framework proceeds in three steps:

\begin{enumerate}
\item \textbf{Adversarial generation.} An attack procedure $\mathcal{A}$ takes the classifier $f_\theta$ and the clean malware features $X$ as input and produces a perturbed set $X_{\text{adv}}=\{x_{i,\text{adv}}\}_{i=1}^m$ intended to be misclassified as benign while respecting a bounded perturbation budget. We consider both unconstrained and similarity-constrained objectives, implemented with iterative FGSM (i-FGSM) and projected gradient descent (PGD).

\item \textbf{Classifier evaluation.} The adversarial examples $X_{\text{adv}}$ are fed into $f_\theta$ to compute the \emph{attack success rate}, defined as the fraction of malware samples classified as benign after perturbation.

\item \textbf{Drift detection.} We compare the distributions of classifier outputs for clean and adversarial malware using drift metrics that quantify the adversarial shift from a distributional perspective.
\end{enumerate}

\begin{figure}[t]
  \centering
  \includegraphics[width=0.8\textwidth]{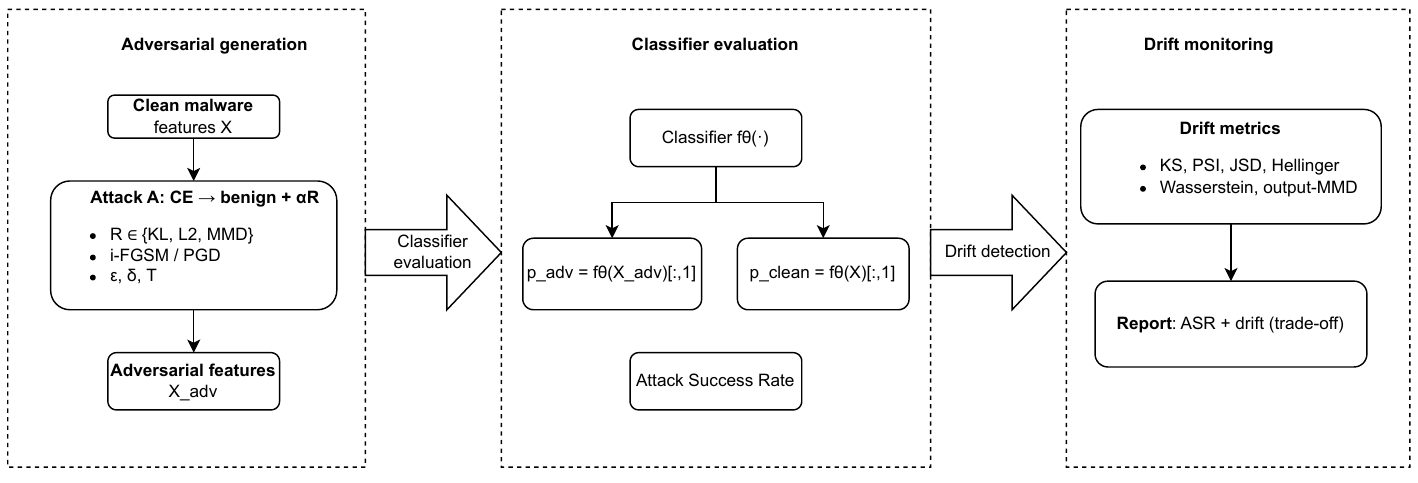}
  \caption{Architecture overview of the methodology components.}
\label{fig:architecture_pipeline}
\end{figure}

\paragraph{Threat Model.}
\label{subsec:threat_model}

We assume a white-box adversary with access to the trained classifier $f_\theta$ and its gradients with respect to the standardized input feature vector $x \in \mathbb{R}^d$. The attacker aims to produce targeted adversarial examples from malware inputs ($y=1$) that are misclassified as benign ($y_{\text{target}}=0$), subject to a per-feature perturbation budget. This models an attacker who can manipulate the observed feature representation within bounded limits.
Our attacks are conducted in the classifier's \emph{feature space}. 
We use feature-space perturbations to quantify the sensitivity of (i) the classifier and (ii) output-based drift monitoring under similarity-aware adversarial objectives. 

\subsection{Adversarial Malware Generation}
\label{subsec:adv_generation}

All attacks are carried out in the standardized feature space of a fixed classifier $f_\theta$. We restrict attention to malware samples ($y=1$) and perform \emph{targeted} attacks toward the benign label ($y_{\text{target}} = 0$). For each malware feature vector $x$, we construct an adversarial counterpart $x_{\text{adv}}$ by minimizing a composite loss under a per-feature perturbation budget.

\paragraph{Adversarial Objective with Similarity Constraints.}
Given a clean malware feature vector $x$ and its adversarial counterpart $x_{\text{adv}}$, we define the general adversarial loss
\begin{equation}
    \mathcal{L}_{\text{adv}}(x_{\text{adv}}) 
    = \mathcal{L}_{\text{CE}}\big(f_\theta(x_{\text{adv}}), y_{\text{target}}\big)
    + \alpha \, \mathcal{R}(x_{\text{adv}}, x, X_{\text{adv}}, X),
    \label{eq:adv_loss_general}
\end{equation}
where $\mathcal{L}_{\text{CE}}$ is the cross-entropy loss encouraging misclassification as benign, $\mathcal{R}$ is an optional similarity regularizer, and $\alpha \ge 0$ controls the strength of the similarity constraint. We consider four choices for $\mathcal{R}$:

\begin{itemize}
 \item \textbf{Baseline (unconstrained):} $\mathcal{R} = 0$, so $\mathcal{L}_{\text{adv}}$ reduces to standard targeted cross-entropy.

    \item \textbf{KL-regularized attack.} We normalize the original and adversarial feature vectors into per-sample probability vectors
    $
        p = \frac{x}{\sum_j x_j + \varepsilon}, 
        q = \frac{x_{\text{adv}}}{\sum_j x_{\text{adv},j} + \varepsilon},
    $
    and define the regularizer
        $\mathcal{R}_{\mathrm{KL}}(x_{\text{adv}},x)
        = \mathrm{KL}(p\|q)$,
    averaged over the batch. 

    \item \textbf{$\ell_2$-regularized attack.} We penalize the Euclidean distance between clean and adversarial features,
    $
        \mathcal{R}_{\ell_2}(x_{\text{adv}},x)
        = \big\|x_{\text{adv}} - x\big\|_2,
    $
    encouraging small perturbations in $\ell_2$ norm.

    \item \textbf{MMD-regularized attack.} We measure discrepancy between batches of clean and adversarial features via the empirical squared MMD (MMD) under a kernel $k$ (linear, RBF, or polynomial). Given a batch of clean features $X$ and adversarial features $X_{\text{adv}}$, we define $
        \mathcal{R}_{\mathrm{MMD}}(X_{\text{adv}},X) = \mathrm{MMD}^2_k(X,X_{\text{adv}}),$
    which encourages the adversarial batch to remain close to the clean malware batch in feature space.
\end{itemize}

\paragraph{Iterative FGSM (i-FGSM).}
\label{subsubsec:ifgsm}
The i-FGSM attack optimizes \eqref{eq:adv_loss_general} via repeated signed gradient updates under a per-feature box constraint. For a clean malware sample $x$, we initialize $x_{\text{adv}}^{(0)} = x$,
and perform $T$ iterations of the form
\begin{equation}
    x_{\text{adv}}^{(t+1)} = 
    \Pi_{[x-\delta,\,x+\delta]}
    \Big(
        x_{\text{adv}}^{(t)} 
        - \epsilon \cdot \operatorname{sign}
        \big(
            \nabla_{x_{\text{adv}}^{(t)}} \mathcal{L}_{\text{adv}}(x_{\text{adv}}^{(t)})
        \big)
    \Big),
    \label{eq:ifgsm_update}
\end{equation}
where $\epsilon$ is the step size, $\delta$ is the perturbation budget (maximum allowed absolute deviation per feature), and $\Pi_{[x-\delta,\,x+\delta]}$ denotes element-wise clipping to the interval $[x_j-\delta,\, x_j+\delta]$ for each feature $j$.

We apply i-FGSM to all four loss configurations. For each configuration and hyperparameter setting $(\epsilon, T, \delta)$, we generate adversarial examples for all malware samples and compute the attack success rate, defined as the fraction of adversarial examples classified as benign by $f_\theta$.

\paragraph{Projected Gradient Descent (PGD) with Random Start.}
\label{subsubsec:pgd}

We also implement a PGD-style attack that uses a \emph{random initialization} within the perturbation set, followed by iterative signed-gradient updates and projection. For a clean sample $x$, we initialize
$
x_{\text{adv}}^{(0)} = \Pi_{[x-\delta,\,x+\delta]}(x + u),  u \sim \mathcal{U}([-\delta,\delta]),
$
and iterate
\begin{equation}
    x_{\text{adv}}^{(t+1)} = 
    \Pi_{[x-\delta,\,x+\delta]}
    \Big(
        x_{\text{adv}}^{(t)} 
        - \eta \cdot 
        \operatorname{sign}
        \big(
            \nabla_{x_{\text{adv}}^{(t)}} \mathcal{L}_{\text{adv}}(x_{\text{adv}}^{(t)})
        \big)
    \Big),
    \label{eq:pgd_update}
\end{equation}
for $t = 0, \dots, T-1$, where $\eta$ is the step size. The random start allows PGD to explore different regions of the feasible set compared to i-FGSM initialized at $x$.

\subsection{Drift Detection Metrics}
\label{subsec:drift_detection}

To quantify how distinguishable adversarial malware is from clean malware, we compare the classifier's output distributions on clean and adversarial samples using a suite of drift metrics. Let $X$ denote the clean malware features from a reference test split and $X_{\text{adv}}$ the corresponding adversarial features produced by a given attack configuration. We compute the predicted malware-class probability for each sample:
$
    p_{\text{clean}} = f_\theta(X)[:,1], 
    p_{\text{adv}}   = f_\theta(X_{\text{adv}})[:,1].
$
We  measure the discrepancy between the one-dimensional distributions of $p_{\text{clean}}$ and $p_{\text{adv}}$ using:
\begin{itemize}
\item \textbf{Population Stability Index (PSI) \cite{khademi2023model}:} a standard feature drift measure, computed by binning $p_{\text{clean}}$ into $B$ bins, using the same bin edges for $p_{\text{adv}}$, and aggregating
    $
        \mathrm{PSI} = \sum_{b=1}^B (c_b - r_b)\log\frac{c_b}{r_b},
    $
    where $r_b$ and $c_b$ are the proportions of clean and adversarial samples in bin $b$.

\item \textbf{Kolmogorov--Smirnov (KS) statistic \cite{massey1951kolmogorov}:} the maximum absolute difference between the empirical cumulative distribution functions (ECDFs) of $p_{\text{clean}}$ and $p_{\text{adv}}$, as in the two-sample KS test.

    \item \textbf{Jensen--Shannon divergence \cite{lin2002divergence}:} a symmetric, bounded divergence between binned probability mass functions derived from $p_{\text{clean}}$ and $p_{\text{adv}}$.

    \item \textbf{Hellinger distance:} 
a metric on probability distributions, computed from the square-rooted bin frequencies of $p_{\text{clean}}$ and $p_{\text{adv}}$. Concretely, if $\hat{p}$ and $\hat{q}$ are the binned empirical distributions, we use
$
    H^2(\hat{p}, \hat{q}) = \frac{1}{2} \sum_b \big(\sqrt{\hat{p}_b} - \sqrt{\hat{q}_b}\big)^2.
$

    \item \textbf{Wasserstein distance \cite{villani2008optimal}:} the one-dimensional Earth Mover's distance between the empirical distributions of $p_{\text{clean}}$ and $p_{\text{adv}}$.

    \item \textbf{MMD \cite{gretton2006kernel}:} an empirical squared MMD between $p_{\text{clean}}$ and $p_{\text{adv}}$ using an RBF kernel with bandwidth chosen by a median heuristic.
\end{itemize}

\section{Experimental Results}
\label{sec:experimental_setup}

We implement the framework described in Section \ref{sec:methodology}  to answer three research questions: (RQ1) How sensitive are output-based drift metrics (JSD, Hellinger, Wasserstein, and output-MMD) compared to KS and PSI under a fixed low-budget operating point? (RQ2)  How do attack design choices  affect the evasion detectability trade-off? (RQ3)  How do attack hyperparameters (step size $\epsilon$, budget $\delta$, and iterations $T$) affect the trade-off between ASR and drift detectability?

\paragraph{Data and Feature Representation.}
\label{subsec:data_features}

We investigate a binary malware detection task using the BODMAS dataset \cite{bodmas}, which comprises 57,293 malware and 77,142 benign Windows PE files. Each sample is represented by a comprehensive 2,381-dimensional feature vector that integrates byte-level statistics and continuous auxiliary features derived from structural and metadata characteristics. To ensure robust analysis, features are standardized using training data statistics and carefully partitioned into distinct training, validation, and test splits.

\paragraph{Classifier Architecture and Training.}
\label{subsec:classifier_setup}
The classifier $f_\theta$ is implemented as a sophisticated feed-forward neural network designed to handle high-dimensional tabular features. The architecture incorporates two fully connected hidden layers with ReLU activations, complemented by batch normalization and dropout techniques to enhance training stability and mitigate overfitting. An $\ell_2$ weight regularization is applied to dense layers, with a softmax output layer generating benign and malware class probabilities. Trained using the Adam optimizer with categorical cross-entropy loss, the model demonstrates exceptional performance, achieving over 95\% accuracy on clean test data.



\paragraph{Adversarial Attack Configurations}
\label{subsec:attack_config}

Our adversarial evaluation framework generates targeted adversarial examples through a comprehensive approach. We explore multiple objectives, including a baseline cross-entropy approach and three sophisticated similarity-constrained regularization techniques: KL divergence, $\ell_2$, and MMD. These objectives are implemented using two advanced optimization methods: iterative FGSM and PGD with random initialization. The attack configurations systematically vary critical hyperparameters such as perturbation budget, step size, iteration count, and regularization strength. Success is quantitatively measured by the proportion of adversarial malware samples misclassified as benign by the target classifier.

\subsection{RQ1: Sensitivity of Drift Metrics Under a Fixed Operating Point}
\label{subsec:rq1}


\begin{table}[t]
\centering
\caption{Drift scores under a fixed operating point ($\epsilon{=}0.0001$, $\delta{=}0.0005$, $T{=}100$, $\lambda_{\mathrm{KL}}{=}\lambda_{\ell_2}{=}\lambda_{\mathrm{MMD}}{=}1$).}
\label{tab:rq1_detectors_fixed}

\small
\setlength{\tabcolsep}{3pt}
\renewcommand{\arraystretch}{1.1}
\begin{adjustbox}{width=0.8\textwidth}
\begin{tabular}{l r r r r r r l}
\toprule
\textbf{Attack} & \textbf{JSD} & \textbf{Hell.} & \textbf{Wass.} & \textbf{MMD} & \textbf{KS} & \textbf{PSI} & \textbf{Detected} \\
\midrule
i-FGSM              & 0.0212 & 0.0176 & 0.000750 & -6.77E-05 & 0.0183 & 0.0025 & No \\
i-FGSM + KL         & 0.0197 & 0.0164 & 0.000550 & -8.15E-05 & 0.0159 & 0.0021 & No \\
i-FGSM + $\ell_2$   & 0.0124 & 0.0103 & 0.000377 & -9.92E-05 & 0.0121 & 0.0008 & No \\
i-FGSM + MMD (Lin)  & 0.1099 & 0.0916 & 0.008141 &  1.80E-03 & 0.0865 & 0.0673 & Yes (MMD) \\
i-FGSM + MMD (RBF)  & 0.1099 & 0.0916 & 0.008259 &  1.81E-03 & 0.0865 & 0.0673 & Yes (MMD) \\
i-FGSM + MMD (Poly) & 0.0565 & 0.0471 & 0.004104 &  6.47E-04 & 0.0453 & 0.0177 & Yes (MMD) \\
\bottomrule
\end{tabular}
\end{adjustbox}
\vspace{-8mm}
\end{table}

Table \ref{tab:rq1_detectors_fixed} reports drift scores computed between the clean and adversarial malware-probability output distributions. Across all statistical metrics, similarity constraints reduce measured drift relative to the unconstrained baseline. For example, the baseline i-FGSM attack yields JSD=0.0212 and Hellinger=0.0176, whereas i-FGSM+KL reduces these to JSD=0.0197 and Hellinger=0.0164, and i-FGSM+$\ell_2$ further reduces them to JSD=0.0124 and Hellinger=0.0103. Consistent reductions are also observed for Wasserstein (0.00075 $\rightarrow$ 0.00055 $\rightarrow$ 0.000377) and for KS statistic (0.0183 $\rightarrow$ 0.0159 $\rightarrow$ 0.0121), indicating that KL and especially $\ell_2$ regularization produce adversarial outputs that are closer to the clean output distribution under all evaluated drift scores.

In contrast, MMD-regularized variants induce substantially larger distribution shifts across every metric. For i-FGSM+MMD with linear and RBF kernels, drift magnitudes increase to JSD=0.1099 and Hellinger=0.0916, with Wasserstein increasing to approximately $8\times10^{-3}$ and the KS statistic increasing to 0.0865. The polynomial-kernel variant yields intermediate shifts (JSD=0.0565, Hellinger=0.0471, Wasserstein=0.004104, KS=0.0453). This consistent separation across statistical metrics indicates that, at this operating point, statistical detectors are sensitive to the stronger output shifts produced by MMD-regularized objectives, while remaining responsive to the smaller shifts induced by KL and $\ell_2$ constraints.

Comparing against traditional metrics, KS aligns with the statistical measures in terms of relative sensitivity: it decreases for KL/$\ell_2$ and increases markedly for MMD-regularized attacks. PSI values remain low across all attacks (0.0008--0.0673), which falls within the negligible-drift range under standard PSI interpretability bands. As a result, PSI is less sensitive than KS and the statistical metrics for distinguishing these attack variants in this low-budget regime.

Under a fixed operating point, statistical drift metrics (JSD, Hellinger, Wasserstein, and MMD) provide a consistent and discriminative signal over similarity-constrained adversarial examples: KL and $\ell_2$ regularization reduce drift magnitudes relative to baseline i-FGSM, whereas MMD-regularized attacks induce substantially larger shifts, are also reflected by KS. PSI remains small for all evaluated attacks, suggesting limited sensitivity in this configuration.

\subsection{RQ2: Effect of Loss Configuration and Optimization Method}
\label{subsec:rq2}

Table \ref{tab:rq2_tradeoff} reports the trade-off between attack effectiveness and drift detectability when varying (i) the optimization procedure (i-FGSM vs.\ PGD with random initialization) and (ii) the attack objective (cross-entropy only vs.\ KL-, $\ell_2$-, and MMD-regularized losses). Unless stated otherwise, all experiments in this subsection use the same perturbation parameters ($\epsilon=0.0001$, per-feature budget $\delta=0.0005$, $T=100$) and regularization strengths ($\lambda_{\mathrm{KL}}=\lambda_{\ell_2}=\lambda_{\mathrm{MMD}}=1$). Drift is quantified on the malware-class output probabilities using KS and PSI, where larger values indicate greater distributional deviation.

\begin{table}[t]
\centering
\caption{ASR and drift trade off across optimization methods and similarity-regularized objectives. All runs use $\epsilon{=}0.0001$, $\delta{=}0.0005$, $T{=}100$, and $\lambda_{\mathrm{KL}}{=}\lambda_{\ell_2}{=}\lambda_{\mathrm{MMD}}{=}1$.}
\label{tab:rq2_tradeoff}

\small
\setlength{\tabcolsep}{4pt}
\renewcommand{\arraystretch}{1.1}
\begin{adjustbox}{width=\textwidth}
\begin{tabular}{l r r r | r r r}
\toprule
& \multicolumn{3}{c|}{\textbf{i-FGSM}} & \multicolumn{3}{c}{\textbf{PGD}} \\
\cmidrule(lr){2-4}\cmidrule(lr){5-7}
\textbf{Objective} & \textbf{Loss} & \textbf{ASR (\%)} & \textbf{KS / PSI}
                  & \textbf{Loss} & \textbf{ASR (\%)} & \textbf{KS / PSI} \\
\midrule
CE              & 6.55556      & 3.10 & 0.0183 / 0.0025 & 6.5555787   & 3.10 & 0.0183 / 0.0025 \\
CE + KL         & 12.568620    & 3.09 & 0.0159 / 0.0021 & 12.451903   & 3.09 & 0.0159 / 0.0021 \\
CE + $\ell_2$   & 6.57777      & 3.09 & 0.0121 / 0.0008 & 6.5903187   & 3.08 & 0.0038 / 0.0001 \\
CE + MMD (Lin)  & 6.237870     & 3.71 & 0.0865 / 0.0673 & 6.55568     & 3.10 & 0.0183 / 0.0025 \\
CE + MMD (RBF)  & 6.228153     & 3.71 & 0.0865 / 0.0673 & 6.555345    & 3.10 & 0.0183 / 0.0025 \\
CE + MMD (Poly) & -1403.280884 & 3.34 & 0.0453 / 0.0177 & -1409.275391 & 3.10 & 0.0159 / 0.0010 \\
\bottomrule
\end{tabular}
\end{adjustbox}
\vspace{-6mm}
\end{table}

\noindent\textbf{Optimizer effect (i-FGSM vs.\ PGD).}
Under the fixed, low-budget setting considered here, replacing i-FGSM with PGD (random start) yields negligible changes in attack effectiveness: ASR remains approximately 3\% across the baseline, KL-, and $\ell_2$-regularized configurations. The drift magnitudes are also similar for these configurations such as baseline KS=0.0183 and PSI=0.0025 for both i-FGSM and PGD, at this perturbation scale, random initialization alone does not materially alter either evasion or output-distribution shift.

\noindent\textbf{Loss configuration effect (CE vs.\ KL/$\ell_2$/MMD).}
The choice of regularization has a clearer influence on drift. Both KL and $\ell_2$ regularization reduce drift relative to the unconstrained objective while maintaining essentially the same ASR. For i-FGSM, KS decreases from 0.0183 (CE) to 0.0159 (CE+KL) and 0.0121 (CE+$\ell_2$), with PSI decreasing from 0.0025 to 0.0021 and 0.0008, respectively. A similar pattern holds for PGD, where $\ell_2$ regularization yields the smallest drift among the evaluated settings (KS=0.0038, PSI=0.0001).

\noindent\textbf{MMD regularization.}
MMD-regularized objectives exhibit behavior that is more sensitive to the kernel choice. For i-FGSM, linear and RBF MMD produce substantially larger shifts (KS=0.0865, PSI=0.0673) while yielding only a modest ASR increase (3.71\%), whereas the polynomial kernel shows smaller drift (KS=0.0453, PSI=0.0177) with ASR close to baseline (3.34\%). For PGD, the tested MMD configurations do not improve ASR and produce drift scores comparable to the baseline/KL settings. Results suggest that, under the present hyperparameters, MMD regularization does not consistently improve the ASR stealth trade-off and may increase detectability depending on the kernel.


\subsection{RQ3: Effect of Attack Hyperparameters}
\label{subsec:rq3}



\begin{table}[t]
\centering
\caption{FGSM-only: representative hyperparameter sweeps illustrating the ASR and drift trade-off.}
\label{tab:rq3_fgsm_sweeps}
\scriptsize
\setlength{\tabcolsep}{3pt}
\renewcommand{\arraystretch}{1.1}
\setlength{\arrayrulewidth}{0.4pt} 

\begin{adjustbox}{width=\textwidth}
\begin{tabular}{l ccccc | ccccc | ccccc}
\toprule
& \multicolumn{5}{c|}{\textbf{Budget sweep} ($\epsilon=0.01$, $T=100$)}
& \multicolumn{5}{c|}{\textbf{Step-size sweep} ($\delta=0.01$, $T=100$)}
& \multicolumn{5}{c}{\textbf{Iterations sweep} ($\epsilon=0.005$, $\delta=0.1$)} \\
\midrule
& 0.005 & 0.01 & 0.02 & 0.05 & 0.1
& 0.001 & 0.005 & 0.01 & 0.02 & 0.1
& 10 & 50 & 100 & 150 & 200 \\
\midrule
$\epsilon$   & 0.01  & 0.01  & 0.01  & 0.01  & 0.01  & 0.001 & 0.005 & 0.01  & 0.02  & 0.1   & 0.005 & 0.005 & 0.005 & 0.005 & 0.005 \\
$\delta$     & 0.005 & 0.01  & 0.02  & 0.05  & 0.1   & 0.01  & 0.01  & 0.01  & 0.01  & 0.01  & 0.1   & 0.1   & 0.1   & 0.1   & 0.1 \\
$T$          & 100   & 100   & 100   & 100   & 100   & 100   & 100   & 100   & 100   & 100   & 10    & 50    & 100   & 150   & 200 \\
ASR (\%)     & 3.10  & 3.50  & 5.24  & 22.37 & 72.04 & 3.50  & 3.50  & 3.50  & 3.49  & 3.49  & 21.71 & 71.76 & 72.04 & 72.04 & 72.04 \\
Avg.\ PSI    & 0.103 & 0.125 & 0.175 & 0.328 & 0.511 & 0.128 & 0.126 & 0.125 & 0.129 & 0.129 & 0.307 & 0.502 & 0.510 & 0.512 & 0.512 \\
\bottomrule
\end{tabular}
\end{adjustbox}
\vspace{-4mm}
\end{table}

\begin{figure}[t]
\centering

\begin{minipage}[t]{0.24\textwidth}
\centering
\includegraphics[width=\linewidth]{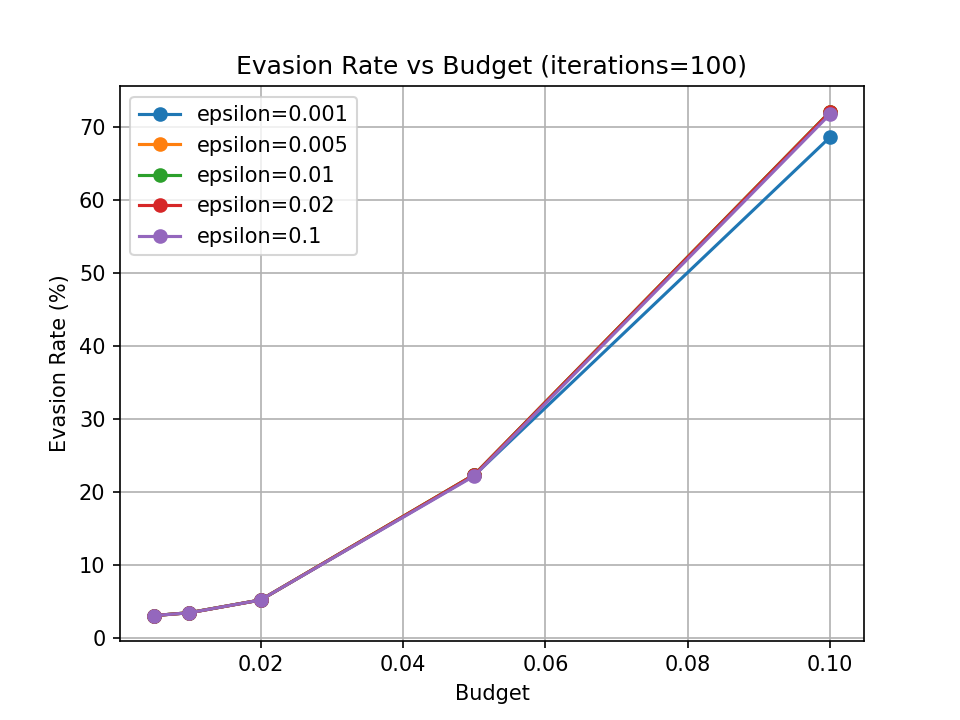}
\captionof{figure}{FGSM ASR vs.\ budget $\delta$.}
\label{fig:fgsm_asr_budget}
\end{minipage}\hfill
\begin{minipage}[t]{0.24\textwidth}
\centering
\includegraphics[width=\linewidth]{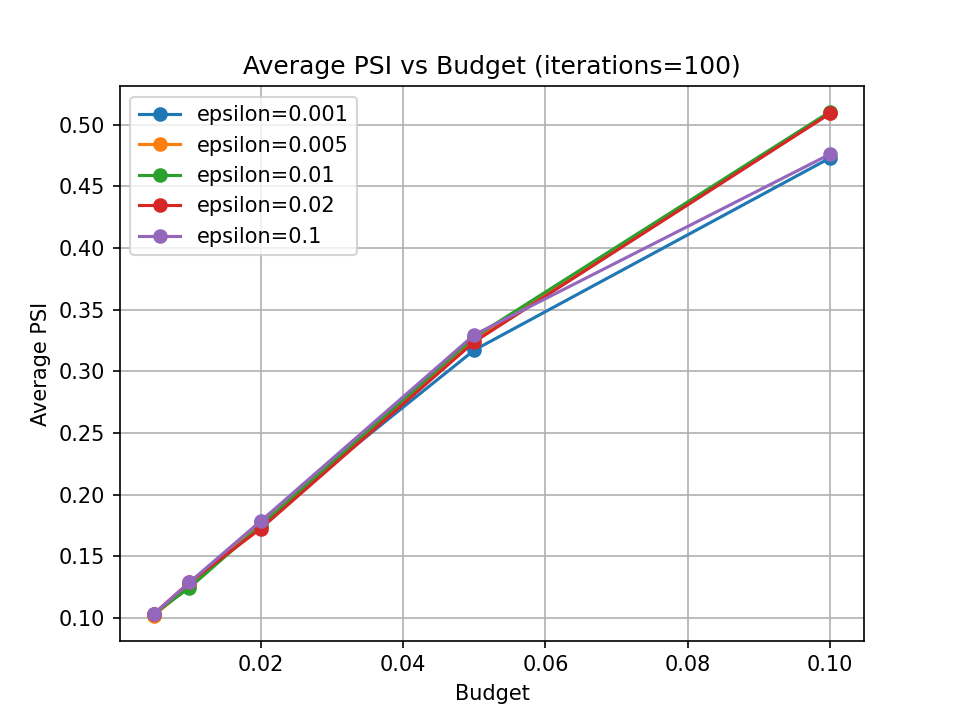}
\captionof{figure}{FGSM Avg.\ PSI vs.\ budget $\delta$.}
\label{fig:fgsm_psi_budget}
\end{minipage}\hfill
\begin{minipage}[t]{0.24\textwidth}
\centering
\includegraphics[width=\linewidth]{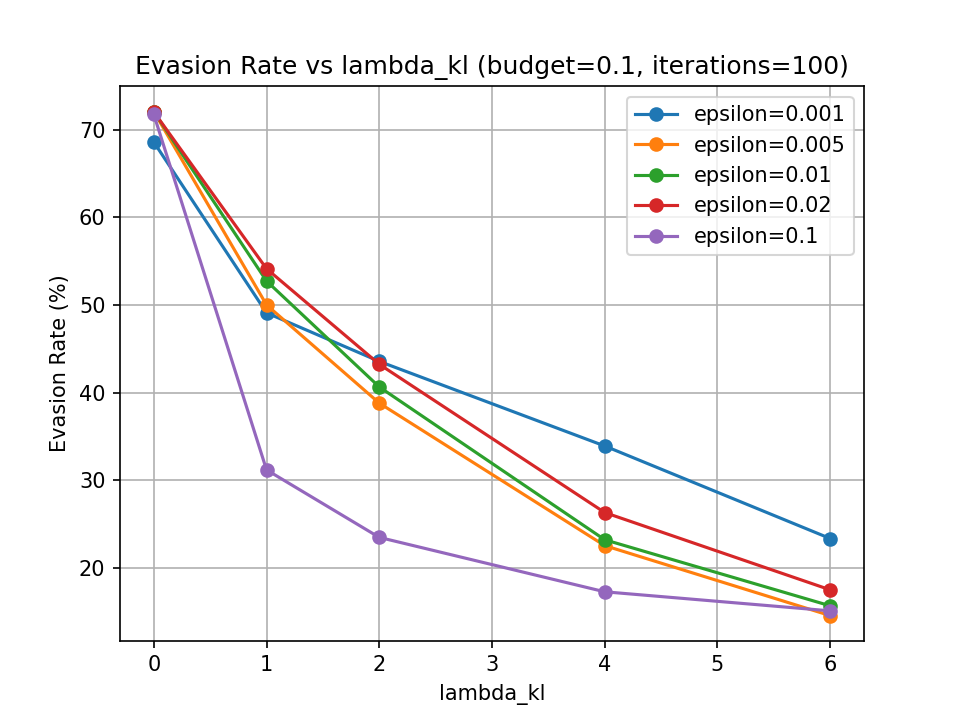}
\captionof{figure}{ASR vs.\ budget $\delta$(KL).}
\label{fig:kl_asr_budget}
\end{minipage}\hfill
\begin{minipage}[t]{0.24\textwidth}
\centering
\includegraphics[width=\linewidth]{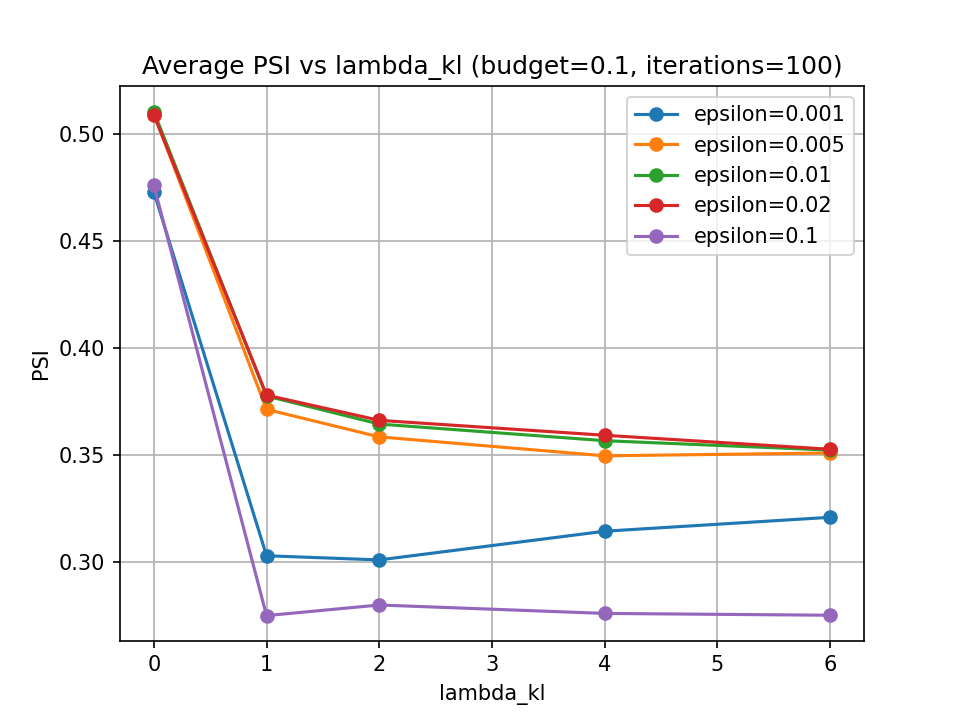}
\captionof{figure}{Avg.\ PSI vs.\ budget $\delta$(KL).}
\label{fig:kl_psi_budget}
\end{minipage}

\end{figure}

\noindent\textbf{FGSM-only (CE).}
As shown in Table \ref{tab:rq3_fgsm_sweeps} and illustrated in Figure \ref{fig:fgsm_asr_budget} and \ref{fig:fgsm_psi_budget}, we  analyze the baseline iterative FGSM attack (targeted CE loss) to understand how the attack hyperparameters such as, step size $\epsilon$, perturbation budget $\delta$, and iteration count $T$ affect the trade-off between attack success rate (ASR) and drift-based detectability (Avg.\ PSI computed on model outputs). 

\noindent\textit{Budget $\delta$ is the dominant driver of the ASR and drift trade-off.}
Increasing the budget produces the largest change in both ASR and drift. Under a representative controlled slice ($\epsilon=0.01$, $T=100$), increasing $\delta$ from $0.005$ to $0.1$ increases ASR from $3.10\%$ to $72.04\%$, while Avg.\ PSI increases from $0.103$ to $0.511$. This pattern reflects the core mechanism of box-constrained FGSM: larger $\delta$ expands the feasible perturbation set $[x-\delta,\,x+\delta]$, making it more likely for the adversarial example to cross the classifier decision boundary, but also inducing larger shifts in the monitored output distribution.

\noindent\textit{Step size $\epsilon$ has a secondary effect once $\delta$ is fixed.}
When the budget is small, varying $\epsilon$ has minimal impact on ASR and Avg.\ PSI because updates quickly saturate under the projection constraint. At $\delta=0.01$ and $T=100$, increasing $\epsilon$ from $0.001$ to $0.1$ leaves ASR essentially unchanged ($3.50\% \rightarrow 3.49\%$), with Avg.\ PSI staying near $0.125$--$0.129$. At larger budgets, $\epsilon$ can mildly modulate the final solution, but its effect remains substantially smaller than the effect of $\delta$.

\noindent\textit{Iteration count $T$ shows diminishing returns and depends on budget.}
At low budgets ($\delta=0.01$), increasing $T$ provides negligible gains because the attack saturates early. At higher budgets, additional iterations can improve ASR initially (for $\delta=0.1$ and $\epsilon=0.005$, ASR increases sharply from $21.71\%$ at $T=10$ to $71.76\%$ at $T=50$), but improvements beyond $T (50$--$100$) are marginal and Avg.\ PSI changes only slightly. $\delta$ primarily controls feasibility (and detectability), while $\epsilon$ and $T$ mainly affect optimization efficiency within the feasible set.

\begin{table}[t]
\centering
\caption{FGSM with KL ($\lambda_{\mathrm{KL}}=1$): representative hyperparameter sweeps illustrating ASR and detectability trends.}
\label{tab:rq3_fgsm_kl_sweeps}
\scriptsize
\setlength{\tabcolsep}{3pt}
\renewcommand{\arraystretch}{1.1}
\setlength{\arrayrulewidth}{0.4pt}

\begin{adjustbox}{width=\textwidth}
\begin{tabular}{l r r r r r | r r r r r | r r r r r}
\toprule
& \multicolumn{5}{c|}{\textbf{Budget sweep} (fix $\epsilon=0.01$, $T=100$)}
& \multicolumn{5}{c|}{\textbf{Step-size sweep} (fix $\delta=0.01$, $T=100$)}
& \multicolumn{5}{c}{\textbf{Iterations sweep} (fix $\epsilon=0.005$, $\delta=0.1$)} \\
\midrule
& $\epsilon$ & $\delta$ & $T$ & \textbf{ASR (\%)} & \textbf{Avg.\ PSI}
& $\epsilon$ & $\delta$ & $T$ & \textbf{ASR (\%)} & \textbf{Avg.\ PSI}
& $\epsilon$ & $\delta$ & $T$ & \textbf{ASR (\%)} & \textbf{Avg.\ PSI} \\
\midrule
1 & 0.01 & \textbf{0.005} & 100 & 2.92 & 0.087  & \textbf{0.001} & 0.01 & 100 & 3.23 & 0.105  & 0.005 & 0.1 & \textbf{10}  & 18.02 & 0.232 \\
2 & 0.01 & \textbf{0.01}  & 100 & 3.23 & 0.098  & \textbf{0.005} & 0.01 & 100 & 3.23 & 0.101  & 0.005 & 0.1 & \textbf{50}  & 50.03 & 0.369 \\
3 & 0.01 & \textbf{0.02}  & 100 & 4.53 & 0.138  & \textbf{0.01}  & 0.01 & 100 & 3.23 & 0.098  & 0.005 & 0.1 & \textbf{100} & 50.01 & 0.371 \\
4 & 0.01 & \textbf{0.05}  & 100 & 17.69& 0.245  & \textbf{0.02}  & 0.01 & 100 & 3.24 & 0.114  & 0.005 & 0.1 & \textbf{150} & 50.01 & 0.371 \\
5 & 0.01 & \textbf{0.1}   & 100 & 52.73& 0.378  & \textbf{0.1}   & 0.01 & 100 & 3.24 & 0.114  & 0.005 & 0.1 & \textbf{200} & 50.01 & 0.371 \\
\bottomrule
\end{tabular}
\end{adjustbox}
\vspace{-4mm}
\end{table}

\noindent\textbf{FGSM with KL regularization.}
We analyze a similarity-constrained FGSM variant where the attack loss augments targeted cross-entropy with a KL penalty. As shown in Table \ref{tab:rq3_fgsm_kl_sweeps} and Figure \ref{fig:kl_asr_budget} and \ref{fig:kl_psi_budget}, we systematically explore how key hyperparameters impact the attack's performance.

\noindent\textit{Budget $\delta$ remains the dominant driver of the ASR and drift trade-off.}
Holding $\epsilon=0.01$ and $T=100$ fixed, increasing the budget from $\delta=0.005$ to $\delta=0.1$ increases ASR from $2.92\%$ to $52.73\%$, while Avg.\ PSI increases from $0.087$ to $0.378$ (Table \ref{tab:rq3_fgsm_kl_sweeps}). As in the baseline FGSM case, the budget directly expands the feasible perturbation set $[x-\delta,\,x+\delta]$, improving the chance of crossing the decision boundary while increasing output-distribution shift.

\noindent\textit{Step size $\epsilon$ has minimal impact in the low-budget regime.}
At $\delta=0.01$ and $T=100$, sweeping $\epsilon$ from $0.001$ to $0.1$ leaves ASR nearly unchanged ($3.23\% \rightarrow 3.24\%$), with only modest PSI variation. The attack saturates under projection, with step size primarily affecting optimization efficiency rather than feasibility.

\noindent\textit{Iterations $T$ yield diminishing returns, especially after early gains at high budget.}
At a larger budget ($\delta=0.1$) and $\epsilon=0.005$, increasing iterations from $T=10$ to $T=50$ substantially increases ASR ($18.02\% \rightarrow 50.03\%$), while Avg.\ PSI also increases ($0.232 \rightarrow 0.369$). Beyond $T\approx 50$, both ASR and Avg.\ PSI largely saturate (Table \ref{tab:rq3_fgsm_kl_sweeps}), indicating diminishing returns from additional iterations once the attack has converged within the feasible set.

\begin{table}[t]
\centering
\caption{FGSM with $\ell_2$ ($\lambda_{\ell_2}=1$): representative hyperparameter sweeps illustrating ASR and detectability trends.}
\label{tab:rq3_fgsm_l2_sweeps}
\scriptsize
\setlength{\tabcolsep}{3pt}
\renewcommand{\arraystretch}{1.1}
\setlength{\arrayrulewidth}{0.4pt}

\begin{adjustbox}{width=\textwidth}
\begin{tabular}{l r r r r r | r r r r r | r r r r r}
\toprule
& \multicolumn{5}{c|}{\textbf{Budget sweep} (fix $\epsilon=0.01$, $T=100$)}
& \multicolumn{5}{c|}{\textbf{Step-size sweep} (fix $\delta=0.01$, $T=100$)}
& \multicolumn{5}{c}{\textbf{Iterations sweep} (fix $\epsilon=0.005$, $\delta=0.20$)} \\
\midrule
& $\epsilon$ & $\delta$ & $T$ & \textbf{ASR (\%)} & \textbf{Avg.\ PSI}
& $\epsilon$ & $\delta$ & $T$ & \textbf{ASR (\%)} & \textbf{Avg.\ PSI}
& $\epsilon$ & $\delta$ & $T$ & \textbf{ASR (\%)} & \textbf{Avg.\ PSI} \\
\midrule
1 & 0.01 & \textbf{0.005} & 100 & 2.61 & 0.048  & \textbf{0.001} & 0.01 & 100 & 2.63 & 0.055  & 0.005 & 0.20 & \textbf{10}  & 2.84 & 0.068 \\
2 & 0.01 & \textbf{0.01}  & 100 & 2.63 & 0.056  & \textbf{0.005} & 0.01 & 100 & 2.63 & 0.056  & 0.005 & 0.20 & \textbf{50}  & 3.51 & 0.090 \\
3 & 0.01 & \textbf{0.02}  & 100 & 2.71 & 0.059  & \textbf{0.01}  & 0.01 & 100 & 2.63 & 0.056  & 0.005 & 0.20 & \textbf{100} & 3.51 & 0.090 \\
4 & 0.01 & \textbf{0.05}  & 100 & 2.85 & 0.067  & \textbf{0.02}  & 0.01 & 100 & 2.62 & 0.052  & 0.005 & 0.20 & \textbf{150} & 3.51 & 0.090 \\
5 & 0.01 & \textbf{0.10}  & 100 & 3.09 & 0.076  & \textbf{0.10}  & 0.01 & 100 & 2.62 & 0.052  & 0.005 & 0.20 & \textbf{200} & 3.51 & 0.090 \\
6 & 0.01 & \textbf{0.20}  & 100 & 3.51 & 0.090  & \textbf{0.20}  & 0.01 & 100 & 2.62 & 0.052  & \multicolumn{5}{c}{ } \\
\bottomrule
\end{tabular}
\end{adjustbox}
\vspace{-4mm}
\end{table}

\begin{figure}[t]
\centering

\begin{minipage}[t]{0.24\textwidth}
\centering
\includegraphics[width=\linewidth]{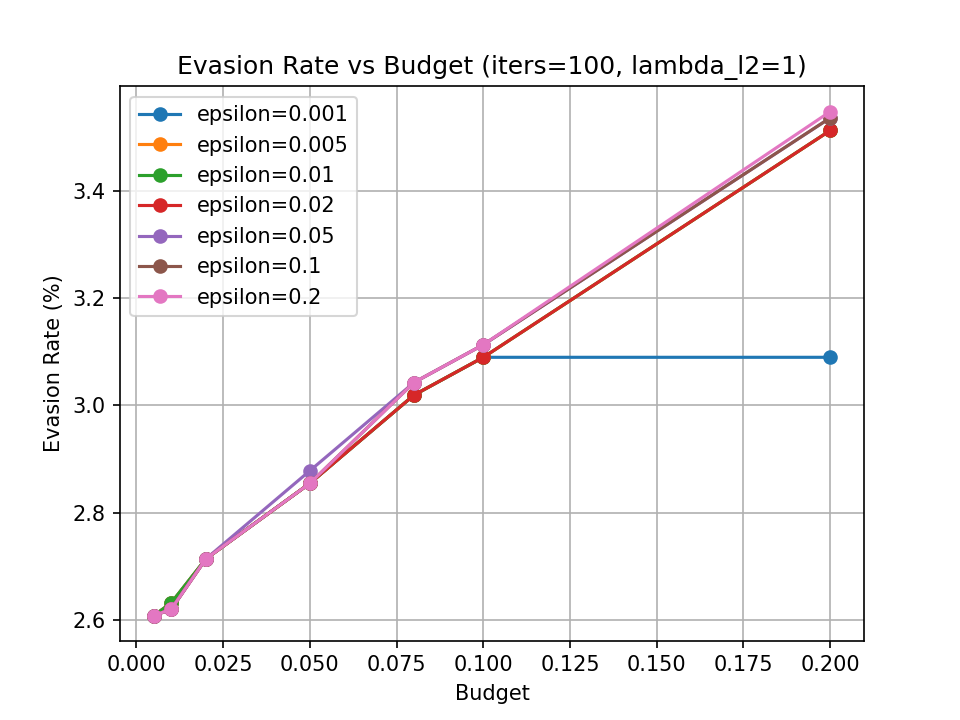}
\captionof{figure}{ASR vs.\ budget $\delta$ ($\lambda_{\ell_2}=1$).}
\label{fig:l2_asr_budget}
\end{minipage}\hfill
\begin{minipage}[t]{0.24\textwidth}
\centering
\includegraphics[width=\linewidth]{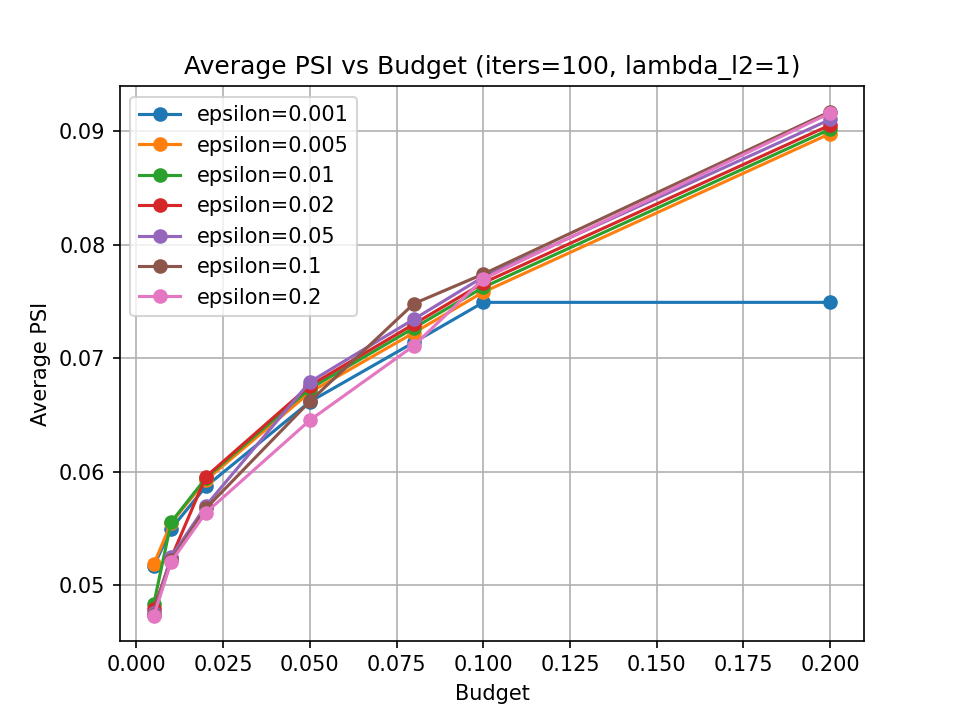}
\captionof{figure}{Avg.\ PSI vs.\ budget $\delta$ ($\lambda_{\ell_2}=1$).}
\label{fig:l2_psi_budget}
\end{minipage}\hfill
\begin{minipage}[t]{0.24\textwidth}
\centering
\includegraphics[width=\linewidth]{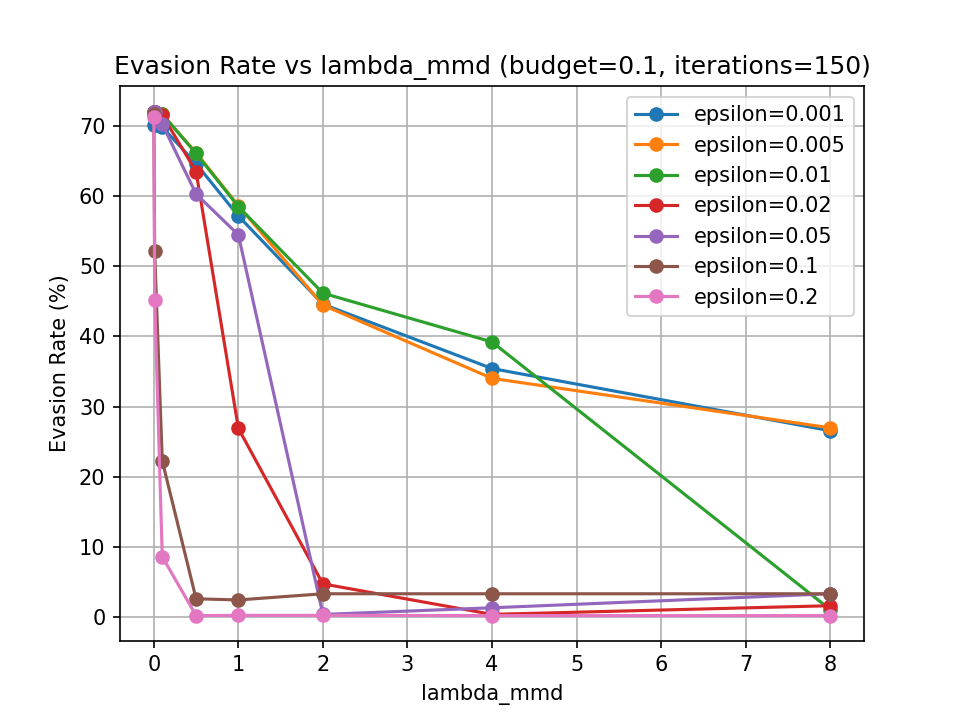}
\captionof{figure}{ASR vs.\ budget $\delta$ ($\delta{\mathrm{MMD}}=0.1$).}
\label{fig:mmd_asr_budget}
\end{minipage}\hfill
\begin{minipage}[t]{0.24\textwidth}
\centering
\includegraphics[width=\linewidth]{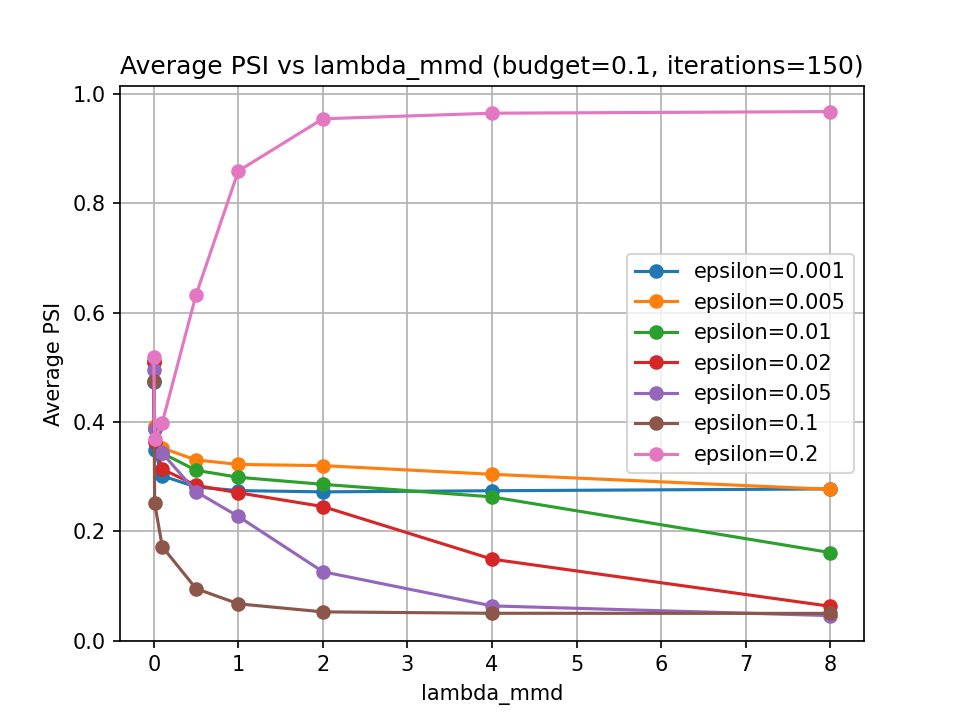}
\captionof{figure}{Avg.\ PSI vs.\ budget $\delta$ ($\delta{\mathrm{MMD}}=0.1$).}
\label{fig:mmd_psi_budget}
\end{minipage}

\end{figure}

\noindent\textbf{FGSM with $\ell_2$ regularization.}
We analyze FGSM augmented with an $\ell_2$ similarity penalty between clean and perturbed feature vectors. As shown in Table \ref{tab:rq3_fgsm_l2_sweeps} and Figure \ref{fig:l2_asr_budget} and \ref{fig:l2_psi_budget}, explore how hyperparameters influence the attack's performance.

\noindent\textit{Budget $\delta$ increases ASR only modestly under $\ell_2$ regularization.}
Unlike the baseline FGSM and FGSM with KL settings, increasing $\delta$ yields only a small improvement in evasion under the $\ell_2$ penalty. Under a representative slice ($\epsilon=0.01$, $T=100$), increasing $\delta$ from $0.005$ to $0.2$ raises ASR from $2.61\%$ to $3.51\%$ (Table \ref{tab:rq3_fgsm_l2_sweeps}). This indicates that the $\ell_2$ penalty strongly constrains the optimization to remain close to the original feature vector, limiting the ability to cross the decision boundary even as the feasible box expands.

\noindent\textit{Step size $\epsilon$ has negligible effect when $\delta$ and $T$ are fixed.}
At $\delta=0.01$ and $T=100$, sweeping $\epsilon$ from $0.001$ to $0.2$ leaves ASR essentially unchanged (approximately $2.62$--$2.63\%$), with only minor changes in Avg.\ PSI (Table \ref{tab:rq3_fgsm_l2_sweeps}). This suggests that, in the low-budget regime, the attack rapidly saturates under the combined effect of projection and the similarity penalty.

\noindent\textit{Iterations $T$ show early gains and then saturate.}
At a larger budget ($\delta=0.2$) and $\epsilon=0.005$, ASR increases from $2.84\%$ at $T=10$ to $3.51\%$ at $T=50$, after which additional iterations yield negligible improvement (Table \ref{tab:rq3_fgsm_l2_sweeps}). Avg.\ PSI exhibits a similar pattern, rising from $0.068$ to $\approx 0.090$ by $T=50$ and then stabilizing.

\begin{table}[t]
\centering
\caption{FGSM with MMD ($\lambda_{\mathrm{MMD}}=1$): representative hyperparameter sweeps illustrating ASR and detectability trends.}
\label{tab:rq3_fgsm_mmd_sweeps}
\scriptsize
\setlength{\tabcolsep}{3pt}
\renewcommand{\arraystretch}{1.1}
\setlength{\arrayrulewidth}{0.4pt}

\begin{adjustbox}{width=\textwidth}
\begin{tabular}{l r r r r r | r r r r r | r r r r r}
\toprule
& \multicolumn{5}{c|}{\textbf{Budget sweep} (fix $\epsilon=0.01$, $T=100$)}
& \multicolumn{5}{c|}{\textbf{Step-size sweep} (fix $\delta=0.01$, $T=100$)}
& \multicolumn{5}{c}{\textbf{Iterations sweep} (fix $\epsilon=0.005$, $\delta=0.20$)} \\
\midrule
& $\epsilon$ & $\delta$ & $T$ & \textbf{ASR (\%)} & \textbf{Avg.\ PSI}
& $\epsilon$ & $\delta$ & $T$ & \textbf{ASR (\%)} & \textbf{Avg.\ PSI}
& $\epsilon$ & $\delta$ & $T$ & \textbf{ASR (\%)} & \textbf{Avg.\ PSI} \\
\midrule
1 & 0.01 & \textbf{0.005} & 100 & 3.10 & 0.096  & \textbf{0.001} & 0.01 & 100 & 3.49 & 0.110  & 0.005 & 0.20 & \textbf{10}  & 19.91 & 0.200 \\
2 & 0.01 & \textbf{0.010} & 100 & 3.48 & 0.108  & \textbf{0.005} & 0.01 & 100 & 3.48 & 0.109  & 0.005 & 0.20 & \textbf{50}  & 92.99 & 0.324 \\
3 & 0.01 & \textbf{0.020} & 100 & 4.99 & 0.140  & \textbf{0.010} & 0.01 & 100 & 3.48 & 0.108  & 0.005 & 0.20 & \textbf{100} & 95.65 & 0.371 \\
4 & 0.01 & \textbf{0.050} & 100 & 17.43& 0.223  & \textbf{0.020} & 0.01 & 100 & 3.48 & 0.111  & 0.005 & 0.20 & \textbf{150} & 95.96 & 0.377 \\
5 & 0.01 & \textbf{0.080} & 100 & 41.55& 0.286  & \textbf{0.050} & 0.01 & 100 & 3.48 & 0.111  & 0.005 & 0.20 & \textbf{200} & 96.04 & 0.379 \\
6 & 0.01 & \textbf{0.100} & 100 & 58.48& 0.299  & \textbf{0.100} & 0.01 & 100 & 3.48 & 0.111  & \multicolumn{5}{c}{ } \\
7 & 0.01 & \textbf{0.200} & 100 & 94.91& 0.333  & \textbf{0.200} & 0.01 & 100 & 3.48 & 0.111  & \multicolumn{5}{c}{ } \\
\bottomrule
\end{tabular}
\end{adjustbox}
\vspace{-4mm}
\end{table}

\noindent\textbf{FGSM with MMD regularization.}
As shown in Table \ref{tab:rq3_fgsm_mmd_sweeps} and Figure \ref{fig:mmd_asr_budget} and \ref{fig:mmd_psi_budget}, we analyze FGSM augmented with an MMD similarity penalty on the targeted cross-entropy objective.

\noindent\textit{Effect of budget $\delta$.}
The perturbation budget produces the largest change in both ASR and Avg.\ PSI. Under a representative slice ($\epsilon=0.01$, $T=100$), increasing $\delta$ from $0.005$ to $0.2$ increases ASR from $3.10\%$ to $94.91\%$, while Avg.\ PSI increases from $0.096$ to $0.333$ (Table \ref{tab:rq3_fgsm_mmd_sweeps}). The budget is the dominant control knob for evasion, while also increasing distributional shift in the monitored outputs.

\noindent\textit{Effect of step size $\epsilon$.}
At fixed $\delta=0.01$ and $T=100$, sweeping $\epsilon$ from $0.001$ to $0.2$ yields negligible changes in ASR (approx. $3.48\%$ throughout) and only small variation in Avg.\ PSI (approx. $0.108$--$0.111$; Table \ref{tab:rq3_fgsm_mmd_sweeps}). In the low-budget regime, step size primarily affects optimization dynamics rather than feasibility.

\noindent\textit{Effect of iterations $T$.}
At larger budget ($\delta=0.2$) and $\epsilon=0.005$, increasing $T$ from $10$ to $50$ increases ASR from $19.91\%$ to $92.99\%$, while Avg.\ PSI increases from $0.200$ to $0.324$. Beyond $T\approx 100$, ASR saturates near $96\%$ while Avg.\ PSI continues to increase slightly (Table \ref{tab:rq3_fgsm_mmd_sweeps}), indicating diminishing returns in evasion and a gradual increase in detectability.

\section{Conclusion}

Our research explored the intricate dynamics of targeted, gradient-based adversarial malware perturbations, focusing on their ability to simultaneously evade deep learning classifiers while maintaining statistical similarity to clean malware across drift metrics.
Similarity constraints using KL divergence and $\ell_2$ distance consistently reduced measured drift across statistical metrics, yet paradoxically did not significantly alter the attack success rate, which remained around 3\%. MMD-regularized variants, in contrast, generated larger output-distribution shifts and proved more detectable.

Our key insights reveal the nuanced relationship between similarity constraints and evasion potential. Constraining perturbations can reduce output-distribution drift without enhancing evasion capabilities, while expanding the perturbation budget substantially increases evasion potential at the cost of increased detectability.
These findings underscore the complex challenge of generating stealthy adversarial malware, emphasizing the critical importance of evaluating attacks through multiple metrics beyond simple misclassification rates. The research illuminates the delicate balance between attack effectiveness and detection probability in the evolving landscape of adversarial machine learning.

\bibliographystyle{splncs04}
\bibliography{references/study_ref}

\end{document}